\RequirePackage{ifpdf}
\documentclass[letterpaper]{JHEP3}
\usepackage{epsfig}
\usepackage{comment}
\usepackage{amsmath}

\def\ba{\begin{eqnarray}}
\def\ea{\end{eqnarray}}
\def\be{\begin{equation}}
\def\ee{\end{equation}}
\def\nn{\nonumber}
\def\exd{{\rm d}}
\def\pd{\partial}
\makeatletter
\def\x@arrow{\DOTSB\Relbar}
\def\xlongequalsignfill@{\arrowfill@\x@arrow\Relbar\x@arrow}
\newcommand{\xlongequal}[2]{%
    \ext@arrow 0099\xlongequalsignfill@{#1}{#2}}
\makeatother

\newcommand{\roughly}[1]{\mathrel{\raise.3ex\hbox{$#1$\kern-0.85em
\lower1ex\hbox{$\sim$}}}}

\newcommand{\lsim}{\roughly<}
\newcommand{\gsim}{\roughly>}

\def\nott#1{\setbox0=\hbox{$#1$}                
   \dimen0=\wd0                                 
   \setbox1=\hbox{/} \dimen1=\wd1               
   \ifdim\dimen0>\dimen1                        
      \rlap{\hbox to \dimen0{\hfil/\hfil}}      
      #1                                        
   \else                                        
      \rlap{\hbox to \dimen1{\hfil$#1$\hfil}}   
      /                                         
   \fi}                                         %

\def\endignore{}
\def\ignore #1\endignore{} 

\def\be{\begin{equation}}
\def\beq\begin{equation}
\def\ee{\end{equation}}
\def\bea{\begin{eqnarray}}
\def\eea{\end{eqnarray}}

\def\eqa{\begin{eqnarray}}
\def\eeqa{\end{eqnarray}}
\def\eq{\begin{equation}}
\def\eeq{\end{equation}}

\def\nn{\nonumber}

\def\pref#1{(\ref{#1})}

\def\ol#1{\overline{#1}}

\def\exd{{\rm d}}
\def\nn{\nonumber}
\def\pref#1{(\ref{#1})}
\def\be{\begin{equation}}
\def\ee{\end{equation}}

\def\beq{\begin{equation}}
\def\eeq{\end{equation}}
\def\beqa{\begin{eqnarray}}
\def\eeqa{\end{eqnarray}}

\def\cA{{\cal A}}

\def\cC{{\cal C}}

\def\cF{{\cal F}}

\def\cL{{\cal L}}

\def\cN{{\cal N}}
\def\cO{{\cal O}}

\def\cR{{\cal R}}

\def\cV{{\cal V}}

\def\ssA{{\scriptscriptstyle A}}
\def\ssB{{\scriptscriptstyle B}}

\def\ssE{{\scriptscriptstyle E}}
\def\ssF{{\scriptscriptstyle F}}

\def\ssL{{\scriptscriptstyle L}}
\def\ssM{{\scriptscriptstyle M}}
\def\ssN{{\scriptscriptstyle N}}
\def\ssP{{\scriptscriptstyle P}}
\def\ssQ{{\scriptscriptstyle Q}}
\def\ssR{{\scriptscriptstyle R}}

\def\ssV{{\scriptscriptstyle V}}

\def\KK{{\scriptscriptstyle KK}}

\newcommand{\bmat}{\left(\begin{array}}
\newcommand{\emat}{\end{array}\right)}

\def\-{\hphantom{-}}

\def\s2{\frac{1}{2}}

\def\Tr{{\rm Tr \,}}

\def\IF{\relax{\rm I\kern-.18em F}}
\def\II{\relax{\rm I\kern-.18em I}}
\def\IP{\relax{\rm I\kern-.18em P}}
\def\IC{\relax{\rm I\kern-.48em C}}
\def\IR{\relax{\rm I\kern-.18em R}}
\def\IK{\relax{\rm I\kern-.20em K}}
\def\IM{\relax{\rm I\kern-.25em M}}

\def\y2{Y_{\ssM\ssN} Y^{\ssM\ssN}}
\def\Riem2{R_{\ssA\ssB\ssM\ssN} R^{\ssA\ssB\ssM\ssN}}
\def\Ricci2{R_{\ssM\ssN} R^{\ssM\ssN}}

\def\f2{F^{a}_{\ssM\ssN} F^{\ssM\ssN}_a}

\def\Asl{\hbox{/\kern-.7500em\it A}} 
\def\dsl{\hbox{/\kern-.5500em$\partial$}}
\def\pxpsl{\hbox{/\kern-.5600em$p$}}
\def\Dsl{\,\raise.15ex\hbox{/}\mkern-13.5mu D}
\def \one{\relax{\rm 1\kern-.26em I}}

\def\exd{{\rm d}}

\def\nn{\nonumber}

\def\({\left(}
\def\){\right)}

\def\Vone{\cV_{1\ssL}}

\def\gR{g_\ssR}
\def\kay{k}

\title{Distributed SUSY Breaking:\\ Dark Energy, Newton's Law and the LHC}

\author{C.P.~Burgess,${}^{1,2}$
L.~van Nierop${}^3$ and
M.~Williams${}^{1,2}$\\

$^1$ Department of Physics \& Astronomy, McMaster University,
 Hamilton ON, Canada\\
$^2$   Perimeter Institute for Theoretical Physics,
 Waterloo ON, Canada\\
$^3$ Abdus Salam ICTP, Strada Costiera 11, Trieste 34014, Italy
}

\date{}

\abstract {We identify the underlying symmetry mechanism that suppresses the low-energy effective 4D cosmological constant within some 6D supergravity models, generically leading to results suppressed by powers of the KK scale, $m_\KK^2$, relative to the much larger size, $m^4$, associated with mass-$m$ particles localized in these models on codimension-2 branes. These models are examples for which the local conditions for unbroken supersymmetry can be satisfied locally everywhere within the extra dimensions, but are obstructed only by global conditions like flux quantization or by the mutual inconsistency of the boundary conditions required at the various branes. Consequently quantities (like vacuum energies) forbidden by supersymmetry cannot become nonzero until wavelengths of order the KK scale are integrated out, since only such long wavelength modes can see the entire space and so `know' that supersymmetry has broken. We verify these arguments by extending earlier rugby-ball calculations of one-loop vacuum energies
within these models to more general pairs of branes within two warped extra dimensions. For the Standard Model confined to one of two otherwise identical branes, the predicted effective 4D vacuum energy density is of order $\rho_{\rm vac} \simeq C (m M_g/4 \pi M_p)^4 = C(5.6 \times 10^{-5} \; \hbox{eV})^4$, where $M_g \gsim 10$ TeV (corresponding to extra-dimensional size $r \lsim 1$ $\mu$m) and $M_p = 2.44 \times 10^{18}$ GeV are the 6D and 4D rationalized Planck scales, and $m$ is the heaviest brane-localized particle. (For numerical purposes we take $m$ to be the top-quark mass and take $M_g$ as small as possible, consistent with energy-loss bounds from supernovae.) $C$ is a constant depending on the details of the bulk spectrum, which could easily be of order $500$ for each of hundreds of fields in the bulk. The value $C \sim 6 \times 10^6$ would give the observed Dark Energy density.
}

\begin{document}

\section{Introduction}
\label{sec:Introduction}

Technically unnatural parameters are those --- {\em e.g.} vacuum energies and scalar masses --- that are measured to be small but which receive large quantum contributions from virtual states at very high energies \cite{tHooft, LesHouchesRev}. They are useful because they provide among the few ways we have to evade general decoupling arguments and acquire a window into what goes on at the very high energies we cannot directly access experimentally.

Supersymmetry is famously useful for naturalness problems because it is among the few symmetries that can forbid vacuum energies and scalar masses, if unbroken. The trick is to design a model that secures the `good' properties (like naturally small vacuum energies or scalar masses) without running into other unacceptable consequences (like super-partners for ordinary particles that are so light they should have already been seen).

Recently, progress has been made on separating these issues within 6D extra-dimensional models \cite{AccSUSY}, with some supersymmetry-breaking effects (like vacuum energies) being naturally at the Kaluza-Klein (KK) scale,\footnote{For extra dimensions that are a two-sphere and fields with KK spectrum $m_\ell^2 = \ell(\ell+1)/r^2$ we take $m_\KK = 1/r$ even though the lowest nonzero KK mass would be $m_1 = \sqrt2/r$.} $m_\KK \simeq 1/r$, even though this is much smaller than the masses, $m$, for the non-supersymmetric particle content on branes (or on a single brane \cite{Teardrop}) localized within the extra dimensions (more about the mechanism for this below). This separation allows the contemplation of realistic models for which the 4D vacuum energy observed in cosmology is technically natural. The models have two (supersymmetric) micron-sized extra dimensions (see below for the origins of this size) setting the scale of the observed Dark Energy density, with ordinary Standard Model particles (but no MSSM superpartners for them) assumed to be
localized on a brane\footnote{See our companion paper \cite{4DSLED} for a take on the long-standing question of what this theory looks like from a four-dimensional perspective.} \cite{Towards, TechNat}.

The existence of such models raises the possibility of performing a meaningful calculation of the vacuum energy, including the contribution of Standard Model particles. This can be done because the dominant contribution now becomes the Casimir energy due to loops of heavy particles in the bulk. And because the UV contributions are suppressed it becomes possible to track precisely how the observed vacuum energy depends on microscopic parameters, to which non-cosmological experiments potentially have access.

In this paper we explore two aspects of such calculations. First we distill out the symmetry mechanism that is at work at the loop level to suppress the size of quantum corrections. We find it to be due to relatively well-known mechanisms combined in a novel way. There are two virtues of formulating the size of the result in terms of symmetry-breaking mechanisms. The first is to clarify whether the same suppressions can be expected also to work for the theory's UV completion (perhaps string theory?) that applies at the highest energies in the bulk. The mechanisms we find at work seem well-suited to arising within string theory.

The second virtue of a symmetry formulation is to clarify the small symmetry-breaking parameters on which the vacuum energy depends, allowing a relatively robust estimate for how the observed dark energy density is related to other scales in the problem, like the higher-dimensional KK and gravity\footnote{For two spherical extra dimensions of radius $r$ our conventions are that $M_p^2 = 4 \pi r^2 M_g^4$, where $M_p := (8\pi G_\ssN)^{-1/2} = 2.44 \times 10^{18}$ GeV is the rationalized 4D Planck Mass. Similarly, the extra-dimensional Planck scale relates to the higher-dimensional Newton constant by $8 \pi G_6 := \kappa^2 := 1/M_g^4$.} scales, $m_\KK$ and $M_g$.
In particular, as shown in more detail below, for the most supersymmetric\footnote{Notice supersymmetry here means supersymmetry of the bulk (or gravity) sector, and does {\em not} mean that superpartners are expected for any brane (or Standard Model) particles.} situations we find the typical contribution of a massive bulk supermultiplet (of string-frame mass $M$) is of order
\be \label{rhovtopwowpreview}
       \rho_{\rm vac}(\hbox{massive mult}) =  \left( \frac{m}{173 \; \hbox{GeV}} \right)^4 \left( \frac{M}{0.1 M_g} \right)^4 \left( \frac{M_g}{10 \; \hbox{TeV}} \right)^4  (5.5 \times 10^{-4} \; \hbox{eV} )^4  \,,
\ee
where $m$ is the mass of the heaviest particle on the brane, which we take for numerical purposes to be the top quark. $M$ cannot be taken as large as $M_g$ without leaving the domain of semiclassical methods ({\em i.e.} the contributions of higher-mass states must be done within a UV completion, such as string theory).

Requiring that formulae like this not be much larger than the observed vacuum energy density,
\be
 \rho_{\rm obs} \simeq (3 \times 10^{-3} \; \hbox{eV})^4 \,,
\ee
puts upper limits on the size of the extra-dimensional gravity scale, and so leads to several testable consequences relating the dark energy density observed in cosmology to scales where other observations might be sensitive. For instance, $M_g \gsim 10$ TeV constrains the expected scales for new physics at the LHC, and implies $r \lsim 1 \; \mu$m, which sets the distance scale at which modifications to Newton's inverse-square force law \cite{Callin} should be visible.

The remainder of this section gives a brief overview of both the origin of the above estimate and the mechanism that underlies the suppression of bulk loops. We start next with the description of the underlying symmetry mechanism (for which more explicit calculations are performed in \S2). This is then followed in the following subsection by a discussion of the parametric dependence of the resulting vacuum energy (with details fleshed out in \S3).

\subsection{SUSY breaking: think globally but act locally}

There are several mechanisms at work that allow the decoupling between the scale, $m$, of particles on the brane and the scale, $m_\KK$, of the energy density observed by cosmology. Before getting to the underlying symmetry mechanism it is worth briefly restating some facts about how vacuum energies arise in 6D models. Part of the story arises already at the classical level, since the classical back-reaction of the bulk geometry to the presence of the branes acts to cancel their tensions in the low-energy 4D theory \cite{Cod2BRflat, Towards}. Since vacuum energy due to loops of brane-localized particles can be regarded as renormalizations of the brane tension, this cancellation of the tension with the back-reaction suggests it should also cancel the influence of vacuum energy loops involving only on-brane particles.

The story is actually a bit more involved than this, with bulk supersymmetry also required to ensure that perturbations of flat solutions also remain flat \cite{LargeDims}. That is, for a non-supersymmetric bulk perturbing the tension, $T \to T + \delta T$, of an initially flat brane configuration turns out to curve the branes by the same amount as would a 4D cosmological constant of size $\delta T$. The classical scale invariance of higher-dimensional supergravity ensures this is not the case for a supersymmetric bulk, however. In the supersymmetric case a sufficient condition for the flatness of the branes turns out to be the absence of a coupling between the brane and a particular bulk field: the six-dimensional dilaton, $\phi$, of the 6D gravity supermultiplet \cite{OtherConical}. In the absence of this coupling the exact (maximally symmetric) solutions to the classical bulk equations give a low-energy 4D world that is precisely flat. (As we see below, the absence of this coupling also has a symmetry
interpretation, inasmuch as its presence would break the supersymmetry of the bulk.)

As a result, the most important quantum corrections are those bulk loops that can induce a brane-$\phi$ coupling \cite{TechNat, AccSUSY}. Ref.~\cite{AccSUSY} shows by explicit calculation how bulk supersymmetry acts to suppress the bulk-loop contributions to the vacuum energy. As is true for many higher-dimensional supergravities, the classical bulk lagrangian can be written in the `string-frame' form \cite{NS}
\be \label{6Ddilatondep}
 \cL_\ssB = e^{-2\phi} L(g_{\ssM\ssN}, \partial_\ssM \phi, \cdots; M) \,,
\ee
where $L$ is a function of the various bulk fields for which $\phi$ appears only differentiated. The dimensions in $\cL_\ssB$ are set by a generic mass scale, $M$, that is of order the higher-dimensional Planck scale (which we shall see to be of order 10 TeV). For later purposes we note that the overall factor of $e^{-2\phi}$ in \pref{6Ddilatondep} shows that $e^{2\phi}$ is the bulk's loop-counting parameter, and its smallness turns out to be related by the field equations to the size of the extra dimensions by a relation of the form $e^\phi \simeq (m_\KK/M)^2$ \cite{SS, ABPQ, Towards}.

Because back-reaction cancels the brane tensions (and so also the direct effects of integrating out brane particles) the dominant contributions to the low-energy effective 4D cosmological constant are obtained by integrating out massive particles in the bulk. Taking these to arise within a lagrangian of the form \pref{6Ddilatondep} gives contributions that are either of order $m^2 m_\KK^2$ or $m_\KK^4$, depending on whether or not the massive states couple directly to the branes, or only couple to them through the intermediary of massless bulk states (like the graviton and its friends in the higher-dimensional supergravity). When the massive states couple directly to the branes their contribution to the vacuum energy is of order $m^2 M^2 e^\phi \simeq m^2 m_\KK^2$. By contrast, when the coupling between branes and massive bulk states only proceeds through a supergravity intermediary\footnote{Actually, since it is only the relevant operators on the brane that give $m^2 m_\KK^2$ contributions \cite{AccSUSY}, for the minimal brane-localized field content it suffices to have the heavy bulk modes not couple directly to the Higgs boson on the brane.} the result instead is of order $M^4 e^{2\phi} \simeq m_\KK^4$.

But what is the symmetry-breaking origin of this story? The reason for the suppression by powers of $m_\KK^2$ can be traced to several mechanisms. These are described in more detail in \S2, but can be stated here in a nutshell:

\medskip\noindent
{\em Bulk Killing Condition:}

\smallskip\noindent
The bulk geometry (in the example of interest, a two-sphere) is such that it does not break one of the higher-dimensional supersymmetries \cite{SS}. That is, the integrability conditions locally allow a nontrivial solution to the Killing spinor equation: $D_\ssM \varepsilon = 0$. Requiring the variation of the dilatino and gaugino also to vanish requires in addition that the background dilaton be constant, $\partial_\ssM \phi = 0$, and that the background magnetic flux quantum be $n = \pm 1$.

\medskip\noindent
{\em BPS Branes:}

\smallskip\noindent
Space-filling 4D branes are situated within the extra dimensions, and in the absence of particles on the branes, the coupling of branes to bulk fields is described by a brane action whose leading terms can be organized in a derivative expansion, whose most general form (unrestricted by supersymmetry) is:
\be \label{eq:TAterms}
 S_b = - \int (T_b \, \omega + \cA_b {}^*F + \cdots) \,,
\ee
where $\omega$ is the brane's volume form and ${}^*F$ is the Hodge dual of the field strength of a particular bulk gauge field, whose nonzero background value plays a role in the flux-stabilization of the extra dimensions \cite{SS}. As usual, the parameter $T_b$ in the no-derivative term describes the brane tension, while the parameter $\cA_b$ of the one-derivative term turns out to describe (see below) the amount of magnetic flux of this bulk gauge field that is localized on the brane \cite{LargeDims}.

This brane action enforces a set of boundary conditions on the bulk fields \cite{HiCoDBCs}, and in general these can obstruct the existence of the Killing spinor and so break the incipient supersymmetry of the bulk. For instance, any coupling of $\phi$ to the brane requires $\phi$ to acquire a nonzero normal derivative near the brane, inconsistent with the vanishing gradient required by bulk supersymmetry. But keeping the first two terms in eq.~\pref{eq:TAterms}, there is only one more condition required for a brane to be consistent with a Killing spinor (beyond the $\phi$-independence of $T_b$ and $\cA_b$). The additional condition requires the coefficients satisfy a `BPS'-like relationship of the form $c_1 T_b + c_2 \cA_b = 0$, for two calculable nonzero quantities $c_1$ and $c_2$. This condition for supersymmetry is BPS-like, in that some supersymmetries can survive the presence of a brane provided its tension and magnetic `charge' are related in a particular way.\footnote{Since $c_1 \ne 0$ supersymmetry is always broken by the brane in the `pure tension' case, with $\cA_b = 0$.} In the example of interest this unbroken supersymmetry is the same one that is preserved by the bulk geometry.

\medskip\noindent
{\em Globally Broken but Locally Unbroken SUSY:}

\smallskip\noindent
What is novel about the `BPS' condition in the example of interest is that the ratio $c_2/c_1$ is a function of $\phi(x_b)$, (again the dilaton, from the bulk gravity multiplet) evaluated at the brane of interest. Furthermore, this field has a classical zero mode, $\phi_0(x) = \varphi \, u_0(x)$, where $u_0(x)$ is a specific normalized mode function and the constant $\varphi$ is undetermined by any of the classical equations of motion (because of a classical scale invariance of the bulk field equations). $\varphi$ turns out ultimately to be fixed by flux quantization in the bulk.

Why is this novel? For each brane there is always a choice for $\phi_b$ for which the BPS condition holds, and preserves supersymmetry, {\em for arbitrary values of $T_b$ and $\cA_b$}. Furthermore, for any one brane this choice is always consistent with the bulk equations of motion because it amounts to a choice for $\varphi$. Of course, if there is more than one brane the same choice for $\varphi$ need not ensure that supersymmetry is preserved at all of them simultaneously. And even for one brane it may happen that flux quantization chooses a value for $\varphi$ that is inconsistent with supersymmetry at the brane.

What is important is that this ensures that any local physics (near a brane or not) doesn't `know' if supersymmetry is broken until it can be determined whether or not the supersymmetric value required for $\phi$ at the brane is consistent with the global configuration of branes and fluxes in the bulk. In particular, quantum effects arising from loops of short-wavelength high-energy modes are local in this way, and since they do not know that supersymmetry is broken these loops cannot generate anything that unbroken supersymmetry forbids. As a result they do not contribute to the low-energy 4D effective vacuum energy as they usually might be expected to do. Instead, a nonzero vacuum energy only arises once wavelengths are integrated out that are large enough to `see' the entire extra dimensions, leading to results that are suppressed by the KK scale, and (if only massless bulk multiplets couple to the brane) are of order $m_\KK^4$.

\subsubsection*{Adding brane particles}

So far this result may not seem very remarkable, since it seems only to say that the SUSY-breaking scale is the KK scale, and so this must be much larger than the observed Dark Energy density. After all, in the above discussion the branes in question did not involve any on-brane degrees of freedom, but if it did one might expect to find superpartner masses at the KK scale.

The remarkable part is that this expectation is wrong \cite{AccSUSY}, ultimately because it makes the mistaken assumption that supersymmetry must be linearly realized on the branes. To investigate this, imagine adding some brane particles and doing so in a way that is not constrained at all by supersymmetry.\footnote{One might wonder how supergravity can couple to such a system, but this can be done by using an equivalent formulation wherein global supersymmetry is nonlinearly realized \cite{NonlinReal} by appropriately coupling a Goldstone fermion on the brane, and then coupling this to supergravity using standard Noether methods. In the end the Goldstone fermion is eaten by the super-Higgs mechanism \cite{SuperHiggs} to give a mass to the massless KK mode of the bulk gravitino.} (For instance one might imagine having a theory of just bosons or just fermions on a brane, or perhaps precisely the Standard Model itself \cite{MSLED}.) In such a picture particle physics on the brane is not supersymmetric at all; indeed it need
not have the particle content to fill out a supermultiplet.

Now comes the main point. Consider integrating out all the brane fields, to determine the low-energy effective cosmological constant. In general all possible effective couplings of the brane to the bulk fields are generated, and none of them need be particularly small. In particular, expanding the result in a derivative expansion again gives eq.~\pref{eq:TAterms}, with renormalized coefficients, $T_b'$ and $\cA_b'$, plus terms involving at least two derivatives. From this point on the discussion proceeds as above, leading for the same reasons to a vacuum energy that is of order the KK scale. It does so because the BPS condition could be satisfied by fixing the value, $\varphi$, of a bulk zero mode, and this works equally well for $T_b'$ and $\cA_b'$ because it took place for {\em any} value of $T_b$ and $\cA_b$.

A similar interplay between locally unbroken but globally broken internal symmetries has also been found to be useful in extra-dimensional versions of Grand Unified theories \cite{LocalGUTS}, and have long been pined for as a potential brane-world mechanism for obtaining a small vacuum energy or scalar mass \cite{OtherGlLo}. Related models also arise within 4D theories, such as with `deconstructed' dimensions \cite{deconstruction} used to produce phenomenological `littlest Higgs' models addressed to the electroweak hierarchy problems of the Higgs boson \cite{Littlest}.

\subsection{The numerology}

In a nutshell, the supersymmetry-breaking story implies (and explicit calculations bear out --- see below and in \cite{AccSUSY}) that for generic brane configurations (assuming no direct brane couplings to $\phi$ or to massive multiplets) the low-energy 4D cosmological constant is of order
\be \label{rhovsummary}
    \rho_{\rm vac} = \frac{C \delta}{(4 \pi r^2)^2} = C \delta \; \left( \frac{M_g^2}{M_p} \right)^4 = C  \left[ \frac{T}{(5 \; \hbox{TeV})^4} \right] \left( \frac{M_g}{10 \; \hbox{TeV}} \right)^4  (0.027 \; \hbox{eV} )^4  \,.
\ee
Here the overall scale is $m_\KK^4$, and $1/(4\pi)^2$ is the generic one-loop factor.\footnote{In 6 dimensions one loop actually brings a factor of $1/(4\pi)^3$, but one of these cancels the $4\pi$ coming from the volume of the extra dimensions.} The factor $\delta \simeq \kappa^2 T/2\pi$ is a dimensionless measure of the size of the brane's gravitational coupling, with $T$ an energy density on the brane. Control of the semiclassical limit requires $\delta \ll 1$, and supersymmetry in the bulk requires $\rho_{\rm vac}$ must vanish as all energy densities on the brane vanish \cite{SS}. Finally, $C$ is a calculable number that is not systematically suppressed by symmetry-breaking parameters.

Clearly, for $C$ order unity, for generic energy densities on the branes eq.~\pref{rhovsummary} cannot be small enough to describe the observed Dark Energy density without also conflicting with existing constraints on extra dimensions. The most important in this case are constraints on astrophysical energy loss in stars and supernovae \cite{Raffelt, SUSYRaffelt, MSLED}, which imply $M_g \gsim 10$ TeV. Taking the lowest value and assuming $T \simeq (5 \; \hbox{TeV})^4$ (motivated by current bounds on the existence at the LHC of `string' excitations of Standard Model particles), we get the numerical values quoted in \pref{rhovsummary}.

The case of identical branes in the extra dimensions turns out to be supersymmetric, and so for this choice the loop contributions to $\rho_{\rm vac}$ vanish. Radiative corrections on the brane (such as would happen if the Standard Model resided on one brane, but not the other) can then make the brane tensions differ from one another, breaking supersymmetry and allowing a nonzero vacuum energy. In this case the brane energy relevant to the suppression factor $\delta$ is this loop-induced tension difference, $\Delta T \simeq \mp m^4/(4\pi)^2$ instead of the overall tension, $T$. Here
the upper (lower) sign assumes the most massive brane particles are bosons (fermions).

This leads to the following, smaller, estimate
\be \label{rhovtop}
  \rho_{\rm vac} = \pm C  \left( \frac{m}{173 \; \hbox{GeV}} \right)^4 \left( \frac{M_g}{10 \; \hbox{TeV}} \right)^4  (1.3 \times 10^{-4} \; \hbox{eV} )^4  \,,
\ee
where for numerical purposes we take the mass of the heaviest known particle --- the top quark --- when evaluating the heaviest brane-particle mass. Depending on the value of $C$, the resulting vacuum energy is now small enough to be consistent with the measured Dark Energy density.

There can be a variety of supermultiplets living in the bulk, including the usual supergravity, gauge and hypermultiplets at the massless level. Indeed, for chiral 6D supergravity \cite{NS} literally hundreds of these multiplets can be required by anomaly-cancellation conditions \cite{6Danomcancel}. However, field-for-field it is the massive multiplets of the bulk supergravity that contribute the most to the vacuum energy, and the value of $C$ for the lowest-spin massive 6D supermultiplet is known, and given in \S3\ as a function of the configuration of branes. In the supersymmetric case of most interest, eq.~\pref{rhovtop}, it is given by \cite{AccSUSY}
\be
  C = \frac12 \, \left( \frac{\kappa M}{2 g_\ssR} \right)^4 \quad \hbox{if $\Delta T < 0$}, \qquad \hbox{and} \qquad  C = \frac12 \, \left( \frac{\kappa M}{2 g_\ssR} \right)^2 \quad \hbox{if $\Delta T > 0$}\,,
\ee
up to corrections that are suppressed by a power of $(2 g_\ssR/\kappa M)^2$. Massless multiplets contribute with coefficients that are of order $(\kappa M/2 g_\ssR)^0$, for which $C$ (and its sign) are unknown. Here $M$ is the mass parameter for the massive multiplet and $g_\ssR$ is the gauge coupling for the flux-stabilizing Maxwell field in the higher-dimensional theory.

Control of approximations in the low-energy theory requires we take $M \lsim M_g$ and $g_\ssR M_g \lsim 1$, and to properly include scales larger than this would require using a UV completion, such as string theory. In this case the result would be obtained by summing expressions like eq.~\pref{rhovtop} over the relevant particle spectrum, and although this calculation cannot yet be done it is tempting to expect that the result is similar, with $M$ replaced by the appropriate string scale. For numerical purposes we can take $M = 0.1 M_g$, $\Delta T<0$ and $g_\ssR M_g = 0.01$ so that $C=250$. This choice gives the estimate quoted in eq.~\pref{rhovtopwowpreview}:
\be \label{rhovtopwow}
  \rho_{\rm vac}(\hbox{massive mult}) = \left( \frac{m}{173 \; \hbox{GeV}} \right)^4 \left( \frac{M_g}{10 \; \hbox{TeV}} \right)^4  (5.5 \times 10^{-4} \; \hbox{eV} )^4  \,.
\ee
It is remarkable that this is smaller in magnitude to the observed value if the largest-mass particle on the brane is the top quark.\footnote{It appears to have the wrong sign if the heaviest particle is a fermion, like the top quark, but because a single low-spin massive multiplet need not dominate the entire bulk result, definitively computing the sign and precise magnitude must await a fuller understanding of the bulk spectrum, potentially including the UV sector. What is remarkable about eq.~\pref{rhovtopwow} is that it is small enough that many fields contributing a similar order of magnitude can give an acceptable value without the need for detailed cancelations.}

This allows us to relate more precisely the scale of the observed Dark Energy density to the other scales in these models, like $r$ (whose value is probed by tests of deviations from Newton's inverse-square law) and extra-dimensional gravity scale, $M_g$ (whose value is relevant to signals in the Large Hadron Collider). We find
\begin{itemize}
\item The extra-dimensional radius is most strongly constrained by the lower limit on $M_g$, since $M_g > 10$ TeV requires $r^2 = M_p^2/(4\pi M_g^4) < (1.4 \; \mu\hbox{m})^2$. This is below, but not excessively far below, the current upper limit, $r_{\rm exp} < 45 \; \mu$m, coming from short-distance tests of Newton's inverse-square law \cite{Adelberger}.
\item Because $\rho_{\rm vac} \propto M_g^4$ it is fairly sensitive to the size of $M_g$. Moving $M_g$ up to 30  TeV would raise $\rho_{\rm vac}$ by a factor of 81. To pin down the precise limits on $M_g$ in this way requires a detailed evaluation of the contribution to $C$ of all bulk supermultiplets, massless and massive, as well as the contribution of states within the UV completion. Although this is beyond the present state of the art, it is clear that once $M_g$ is too large detailed cancelations would be required in order to produce an acceptably small result. For instance, given the hundreds of multiplets likely to be present in the bulk a conservative guess might take the complete result for $\rho_{\rm vac}$ to be $1000 \simeq 5.6^4$ times larger than in eq.~\pref{rhovtopwow}, in which case there is little room to allow $M_g$ to be much larger than 10 TeV. If the full result instead were only 100 times the size of \pref{rhovtopwow} then $M_g$ could be allowed to be as large as 20 TeV. (Fig.~\ref{ds02branes} shows graphically the range of allowed values of $M_g$ and $r$ allowed by the above formulae.)
\item The size of $M_g$ is relevant for the LHC since any new states associated with the UV completion (such as string excitations of Standard Model particles) are likely to have masses {\em below} $M_g$. For instance string states generally have masses that are smaller than the gravity scale by powers of the string constant, so in any weakly coupled string theory states would be expected below $M_g$. Although precise constraints would require a better understanding of the phenomenology of weak-scale strings \cite{WSStrings}, an indication of the likely strength of these bounds can be found from limits placed on KK excitations of Standard Model particles in more conventional extra-dimensional models, and are already at several TeV \cite{LHCBounds}.
\end{itemize}

We now turn to a more detailed derivation of these results.

\section{Distributed SUSY breaking}

We start with the discussion of the bulk system, and the local and global conditions for unbroken supersymmetry.

\subsection{The system of interest}

The bulk theory we explore explicitly is 6D chiral gauged supergravity \cite{NS}, with non-trivial background fields taken to be the graviton $g_{\ssM\ssN}$, a gauge field, $F_{\ssM\ssN}$, and the dilaton $\phi$. Their equations of motion follow from the action\footnote{We use a `mostly-plus' metric, and adopt Weinberg's curvature conventions \cite{GandC}.}
\be
 S = -\int \! \exd^6 x \sqrt{-g} \bigg[ \frac1{2\kappa^2} \Big( \cR + \pd_\ssM \phi \, \pd^\ssM\phi \Big) + \frac{e^{-\phi}}{4 \gR^2} \, F_{\ssM\ssN} \, F^{\ssM\ssN} + \frac{2 \gR^2}{\kappa^4} \, e^\phi \bigg] \,;
\ee
and are explicitly given by
\begin{gather} \label{NSFEs}
 \cR_{\ssM\ssN} + \pd_\ssM\phi\,\pd_\ssN\phi + \frac{\kappa^2 e^{-\phi}}{\gR^2} F_{\ssM\ssP} {F_\ssN}^\ssP - \frac12\left( \frac{\kappa^2e^{-\phi}}{4 \gR^2} F_{\ssP\ssQ} F^{\ssP\ssQ} - \frac{2\gR^2}{\kappa^2} \; e^\phi \right) g_{\ssM\ssN} = 0 \qquad\\
 \pd_\ssM \Big(\sqrt{-g} e^{-\phi} F^{\ssM\ssN} \Big) = 0 \\
 \square \phi + \left( \frac{\kappa^2e^{-\phi}}{4\gR^2} F_{\ssP\ssQ} F^{\ssP\ssQ} - \frac{2\gR^2}{\kappa^2} \; e^\phi \right) = 0 \,.
\end{gather}
The gauge field appearing here gauges a specific $R$-symmetry, $U(1)_\ssR$, of the 6D supersymmetry algebra. Notice these equations are invariant under the classical rigid scaling symmetry under which $g_{\ssM\ssN} \to \zeta \, g_{\ssM \ssN}$ and $e^{-\phi} \to \zeta e^{-\phi}$.

\subsection{Local and global conditions for SUSY}

Our interest is in when the solutions to these equations preserve an unbroken supersymmetry. We first identify the necessary conditions that must hold locally within the bulk (both near and far from the branes), and then ask whether there are global obstructions to extending these conditions due to the existence of boundary conditions at the positions of branes or flux quantization.

\subsubsection*{Local conditions deep in the bulk}

A configuration of the given background fields is supersymmetric if there exists a nonzero supersymmetry parameter, $\epsilon$, for which the following transformations vanish once evaluated at the background:\footnote{In general the dilatino transformation also contains a term involving a bulk Kalb-Ramond field, but this vanishes in the geometries of interest here.}
\begin{gather}
 \delta\lambda = \frac1{2\sqrt 2 \, \gR} \, e^{-\phi/2} F_{\ssM\ssN} \Gamma^{\ssM\ssN} \epsilon - \frac {i\sqrt 2 \, \gR}{\kappa^2} \, e^{\phi/2} \epsilon \nn\\
 \delta\chi = \frac1{\kappa\sqrt 2} (\pd_\ssM\phi) \Gamma^\ssM \epsilon \\
 \delta\psi_\ssM = \frac{\sqrt 2}\kappa \, D_\ssM \epsilon \nn\,,
\end{gather}
where $\lambda$, $\chi$ and $\psi_\ssM$ are respectively the 6D gaugino, dilatino and gravitino that partner with the nontrivial background fields.

First consider the dilatino condition, $\delta \chi = 0$, which implies the dilaton must be a constant: $\partial_\ssM \phi = 0$. Solutions with constant $\phi$ are possible \cite{SS, Towards}, and as we see below it requires the solution locally to have the spherical rugby-ball form.

Next, the gaugino condition $\delta \lambda = 0$ can also allow nonzero $\epsilon$, provided its decomposition in terms of 4D spinors has the form
\be \label{E:6DWeylspinor}
 \epsilon = \left(
 \begin{array}{c}
 \varepsilon_{4\pm} \\  0
 \end{array}
 \right) \,,
\ee
where the 4D spinor $\varepsilon_{4\pm}$ satisfies the 4D Weyl condition $\gamma_5 \varepsilon_{4\pm} = \pm \varepsilon_{4\pm}$, with the sign correlated with that of the flux-quantization integer, $n = \pm 1$.

Finally the condition $\delta \psi_\ssM = 0$ boils down to the existence of a covariantly constant (Killing) spinor:
\be
 D_\ssM\epsilon= \left(\pd_\ssM - \frac{i}4 \Gamma_{\ssA\ssB} \, \Omega_\ssM^{\ssA\ssB} - i  A_\ssM \right)   \epsilon  = 0 \,,
\ee
where the covariant derivative of $\epsilon$ depends on $A_\ssM$ because the corresponding symmetry is an $R$ symmetry (and so does not commute with supersymmetry). The integrability condition for such a spinor states $[D_\ssM , D_\ssN] \epsilon = -i  \left( \frac12 \, R_{\ssM \ssN \ssP \ssQ} \Gamma^{\ssP \ssQ} + F_{\ssM \ssN} \right) \epsilon = 0$, which for the rugby-ball backgrounds is automatically satisfied by eq.~\pref{E:6DWeylspinor} together with the 4D condition $\gamma_5 \varepsilon_{4\pm} = \pm \varepsilon_{4\pm}$. The resulting Killing spinor turns out to be a constant on a patch in the bulk, due to a cancellation in the spinor covariant derivative between the $R$-symmetry gauge connection, $A_\varphi$, and the rugby-ball spin connection, $\Omega^{45}_\varphi$.

It is in this way that we see that nontrivial solutions to the Killing-spinor equation exist on any local coordinate patch, and what remains is to see if these solutions can be stitched together to satisfy all of the boundary conditions set by the problem around the extra dimensions. Ref.~\cite{SS} shows how this can be done to assemble a global Killing spinor if the extra-dimensional metric is a sphere, showing that the Salam-Sezgin spherical solution preserves one 4D bulk supersymmetry.

\subsubsection*{Local conditions near a brane}

Once branes are present we can ask whether they can obstruct the local existence of supersymmetric patches that include the brane position. In general they do, by dictating near-brane boundary conditions that are not satisfied by any solutions to the conditions for unbroken supersymmetry. One way this can happen in the present instance would be for the brane actions not to be stationary with respect to variations of the 6D dilaton, $\phi$. This would preclude the existence of supersymmetric configuration in a patch including the brane because back-reaction makes $\delta S_b/\delta \phi$ proportional to the near-brane limit of $\rho \, \partial_\rho \phi$ (where $\rho$ is the radial proper distance from the brane). Consequently, having a nonzero $\delta S_b/\delta \phi$ would contradict the requirement found earlier that $\phi$ be a constant throughout the patch. A sufficient condition for this obstruction not to arise is to have all of the coefficient functions, $T_b$, $\cA_b$ {\em etc.}, be completely independent of $\phi$.

But branes can also break supersymmetry even if they do not couple to the dilaton. They can do so because of the changes the brane-localized flux interaction implies for the near-brane boundary conditions for the bulk gauge field, $A_\ssM$. The background gauge potential on a patch near a brane satisfying the near-brane boundary conditions dictated by back-reaction \cite{LargeDims} turns out to be given by
\be
  A_\varphi = -\frac{\cN\alpha}2 (\cos\theta-b) + b \, \Phi_b \,,
\ee
where $\cN$ and $b$ are signs, $\pm 1$, with $\cN$ set by the flux quantization integer and $b$ labeling the two branes situated at $\cos \theta = b$. (In what follows, we'll write the $\cN$ explicitly as $\pm$, but leave $b$ intact.) The non-trivial component of the spinor covariant derivative then becomes
\be
 D_\varphi \epsilon = \left[\pd_\varphi -\frac{i}2 \left(
 \begin{array}{cc}
 \gamma_5 & 0 \\
 0 & -\gamma_5
 \end{array}
 \right) (\alpha \cos\theta -b) \pm \frac{i\alpha}2 (\cos\theta-b) - ib \, \Phi_b  \right] \epsilon =0\,,
\ee
and so in a patch near a brane $\varepsilon_{4\pm}$ must satisfy
\be
 \left\{ \partial_\varphi + ib \left[ \pm \frac12 (1 - \alpha) - \Phi_b \right] \right\} \varepsilon_{4\pm} = 0 \,.
\ee
This can have nontrivial solutions if the brane defect angle, $\alpha_b$, and flux, $\Phi_b$, are related by
\be \label{eq:branesusycond}
 \pm \frac12 (1 - \alpha_b) = \Phi_b \,.
\ee

\subsubsection*{Global obstructions}

We now ask whether these local conditions for unbroken supersymmetry can be assembled together to give a global solution that respects all boundary conditions. Since we know this can be done when branes are absent \cite{SS}, it suffices to check whether the various near-brane boundary conditions -- like eq.~\pref{eq:branesusycond} -- can be consistent with one another, and with other global conditions like flux quantization.

As was shown in \cite{AccSUSY}, a single 4D supersymmetry can survive all these conditions when the two branes are identical --- {\em i.e.} have equal tensions and localized fluxes --- and do not couple to the dilaton, $\phi$. This can partially be seen from the consistency of the above local conditions for supersymmetry near each brane, eq.~\pref{eq:branesusycond}. Furthermore, these conditions turn out to be consistent with flux quantization, which for identical branes turns out to require \cite{AccSUSY}
\be \label{E:SUSYcond}
 \Phi_+ = \Phi_- = \frac{\Phi}{2} = \pm \frac12\, (1 - \alpha) = \pm \frac{\delta}{4\pi} \,,
\ee
where $\Phi := \Phi_+ + \Phi_-$ defines the total localized flux.

In general, though, the branes break supersymmetry. This is true in particular for `pure-tension' branes, for which $\Phi_b = 0$. For such branes any nonzero brane tension --- $\alpha \ne 1$ --- necessarily breaks supersymmetry. As argued in the introduction, local UV physics that sees only one brane doesn't know supersymmetry breaks if this is only due to an inconsistency between the properties of different branes.

\subsection{Explicit solutions for non-identical branes}

It is useful to make the above considerations concrete by presenting the explicit solutions to the bulk field equations appropriate to a generic pair of branes. The only assumption these solutions make is that neither brane couples directly to the bulk dilaton, $\phi$, and as a result the normal derivative of $\phi$ vanishes in the near-brane limit \cite{HiCoDBCs}. The solutions described here are those of refs.~\cite{GGP}, written in a more physically transparent coordinate system. (See Appendix \ref{app:GGPcoords} for the explicit relationship with the forms given in \cite{GGP}.)

The metric which solves eqs.~\pref{NSFEs} has the form
\be \label{eq:GGPmetric}
 \exd s^2 = W^2(\theta) \, \exd s_4^2 + r^2(\theta) \Big(\exd\theta^2 + \alpha^2(\theta) \sin^2\theta \,\exd\varphi^2 \Big) \,,
\ee
where $\exd s_4^2$ denotes a maximally symmetric four-dimensional geometry, $\exd s_4^2 = \hat g_{\mu\nu} \,\exd x^\mu \exd x^\nu$, and the field equations imply $\hat g_{\mu\nu} = \eta_{\mu\nu}$ and
\be
 r(\theta) =  r_0 W(\theta) \qquad \hbox{with} \qquad r_0 := \frac{\kappa\,e^{-\phi_0/2}}{2\gR} \,,
\ee
and so
\be
 \exd s^2 = W^2(\theta) \Big[ \exd s_4^2 + r_0^2 \Big(\exd\theta^2 + \alpha^2(\theta) \sin^2\theta \,\exd\varphi^2 \Big)\Big] \,.
\ee
The remaining metric functions are
\be
 \alpha(\theta) = \frac\lambda{W^4(\theta)} \,,
\ee
and
\bea
 W^4(\theta)  &=& e^{\xi} \sin^2\frac{\theta}{2} + e^{-\xi} \cos^2 \frac{\theta}2 \nn\\
 &=& \cosh \xi - \sinh \xi \,\cos\theta \,.
\eea
The background gauge field is given by
\be
 F_{\theta\varphi} = \pm \frac{\lambda \,\sin\theta}{2 \, W^8(\theta)} = \pm\frac1{2 \, r^2(\theta)} \, \frac{\epsilon_{\theta\varphi}}{W^4(\theta)} \,,
\ee
where $\epsilon_{mn}$ is the extra-dimensional Levi-Civita tensor. The dilaton is similarly given by
\be
 e^{\phi(\theta)} = \frac{e^{\phi_0}}{W^2(\theta)} \,.
\ee

Here $\xi$, $\lambda$ and $\phi_0$ are three integration constants that can be related to brane properties by the near-brane boundary conditions \cite{HiCoDBCs}. Two of them ($\xi$ and $\lambda$) can be traded for the defect angles, $\delta_b = 2\pi(1 - \alpha_b)$, due to the branes located at the two poles. Our notation writes $\alpha_+ := \alpha(\theta = 0)$, $\alpha_- := \alpha(\theta = \pi)$, $W_+ := W(\theta = 0)$ and $W_- := W(\theta = \pi)$, and so we have
\be
 \alpha_b = \frac\lambda{W_b^4} = \lambda \, e^{b\xi}\,,
\ee
and so
\be \label{eq:Wpmforms}
 \lambda = \sqrt{\alpha_+\alpha_-} \qquad \hbox{and} \qquad e^{\xi} = \sqrt{\frac{\alpha_+}{\alpha_-}} = W_-^4 = \frac{1}{W_+^4} \,.
\ee
In terms of these $\alpha(\theta)$ is given simply by
\be
 \frac1{\alpha(\theta)} = \sum_b\frac1{\alpha_b}\left(\frac{1+b\cos\theta}2\right)
 = \frac{1}{\alpha_+} \; \cos^2 \frac{\theta}2 + \frac{1}{\alpha_-} \; \sin^2\frac{\theta}{2} \,,
\ee
and
\be
 W^4(\theta) = W_+^4 \cos^2 \frac{\theta}2 + W_-^4 \sin^2\frac{\theta}{2} \,.
\ee
In particular, in the special case $W_+ = W_-$ the function $W(\theta)$ (and so also $\phi(\theta)$, $r(\theta)$ and $\alpha(\theta)$) becomes constant, and the geometry \pref{eq:GGPmetric} reduces to the simple rugby-ball solution \cite{SS, Towards}.

\subsection*{Flux quantization}

The third integration constant, $\phi_0$, is completely unfixed by the bulk equations of motion, because of their invariance under constant scale transformations. The condition that ultimately fixes $\phi_0$ is instead flux quantization.

For the systems of interest it is important that the branes carry localized tubes of the background flux themselves \cite{LargeDims}, as in eq.~\pref{eq:TAterms}. In terms of the coefficients in this lagrangian the localized flux contributions on each brane is given by
\be
\Phi_b = \frac{\cA_b e^{\phi_b}}{2\pi} = \frac{\cA_b \,e^{\phi_0}}{2\pi W_b^2}  \,,
\ee
where $\phi_b$ denotes $\phi$ evaluated at the corresponding brane. Consequently the flux quantization condition (for flux quantum $n = \pm 1$) ensures the otherwise-unspecified zero-mode $\phi_0$ adjusts to satisfy
\be
\pm 1 = \sum_b \Phi_b + \frac1{2\pi} \! \int F \,.
\ee
More explicitly, using
\be
 \int F = \int \!\exd\theta \exd \varphi \,F_{\theta\varphi} = \pm 2\pi \lambda =\pm 2\pi \sqrt{\alpha_+\alpha_-} \,,
\ee
together with eq.~\pref{eq:Wpmforms}, we find that the zero-mode is given by
\be
 e^{\phi_0} = \pm \frac{2\pi\Big(1-\sqrt{\alpha_+\alpha_-}\Big)}{\sum_b \cA_b  \left(\frac{\alpha_+}{\alpha_-}\right)^{b/4}} \,.
\ee

\subsection*{Spin and gauge connections}

There are several reasons why this configuration breaks supersymmetry (unless $W_+ = W_-$). First, it does so because $\phi$ generically has a nontrivial gradient, $\partial_m \phi \ne 0$. Second, the gauge and spin connections in general cannot be identified. To see this, note that the gauge field found by integrating the field strength (starting from the flux-localized boundary condition, $A_\varphi(\theta_b) = b \, \Phi_b$ \cite{LargeDims}) is
\be
 A_\varphi(\theta) = b\,\Phi_b \pm \frac\lambda 2 \int_{\theta_b}^\theta \!\exd\theta' \,\frac{\sin\theta'}{W^8(\theta')} = b \,\Phi_b \mp \frac{\alpha_b}{2\,W^4(\theta)} (\cos\theta-b ) \,.
\ee
By contrast, the extra-dimensional component of the spin connection evaluates to
\be
 {\Omega_\varphi}^{45} = \alpha\left( \cos\theta - \frac{3\sin^2\theta}{4 W^4} \, \sinh \xi \right) -b \,,
\ee
and so these connections cannot cancel in the Killing spinor equation, except at the position of the branes provided the supersymmetry condition there,
\be
 \Phi_b = \pm \frac12 (1-\alpha_b) \,,
\ee
is satisfied.

\section{4D vacuum energy and scales}

We now give a quick review of the 1-loop computation of the vacuum energy described for rugby ball solutions in \cite{AccSUSY} and use it to estimate the contributions due to a brane loop in the more general spacetimes sourced by branes that are not identical.

We would like to obtain an effective potential due to loops of various fields in a massive supermultiplet, computed on the classical background described previously. As it turns out, the 1PI effective potential in the case of a warped geometry can be inferred from the rugby-ball result obtained in \cite{AccSUSY} (more on this later). We begin with a brief summary of the methods used and results obtained in \cite{AccSUSY}, before extending them to the warped case of interest here, for which estimates are made.

\subsection{Mode Sums and Renormalization}

We wish to compute the change to the 4D vacuum energy due to a loop of various particles in a massive multiplet. To this end we consider the 1PI quantum action, $\Gamma = S + \Sigma$, where
\be
\Sigma = - \int \! \exd^4 x \,\Vone = \frac{i}2 (-)^\ssF \Tr {\rm Log} \left(\frac{-\square_6 + X+m^2}{\mu^2}\right) \,.
\ee
($X$ denotes additional operators specific to the type of field in the loop; bosons/fermions contribute with $(-)^\ssF = \pm1$.) Wick rotating to Euclidean signature and performing a heat-kernel expansion \cite{GilkeydeWitt, GdWrev}, we have
\eqa
    \Vone &=& \frac12 \, (-)^\ssF \, \mu^{4-d} \sum_{jn} \int \frac{\exd^d k_\ssE}{(2\pi)^d} \, \ln \left( \frac{k_\ssE^2 + m^2 + m_{jn}^2}{\mu^2}  \right) \nn\\
    &=& -\frac{\mu^{4-d}}{2(4 \pi r^2)^{d/2}}  \int_0^\infty \frac{\exd t}{t^{1 + d/2}} \, e^{- t (m r)^2} \, S(t) \,,
\eea
where $m_{jn}^2 = \lambda_{jn}/r^2$ denote the eigenvalues of $-\Box_2 + X$ in the compactified space,
\be
r:=\frac{\kappa \, e^{-\phi/2}}{2\gR} \,,
\ee
and $d = 4 - 2 \, \varepsilon$ with regularization parameter, $\varepsilon$, taken to zero after all divergences in this limit are renormalized. The function $S(t)$ is defined by
\bea
 S(t) &:=& (-)^\ssF \sum_{jn} \exp \left[ - t \lambda_{jn}  \right]
\eea
and has the following small-$t$ expansion:
\bea
S(t) &\simeq&  \frac{s_{-1}}{t} +
 \frac{s_{-1/2}}{\sqrt t} + s_0 + s_{1/2}\, \sqrt{t} \nn\\
 &&\quad  + s_1 \, t + s_{3/2}\, t^{3/2} + s_2 \, t^2 + \cO(t^{5/2})\,.\quad
\eea
Its small-$t$ limit is of interest because it is only a few of the first terms in this series that contribute to the UV divergences appearing in $\Vone$:
\be \label{eq:Vinfty}
 \Vone = \frac{\cC}{(4\pi r^2)^2} \left[ \frac1{4-d} + \ln\left(\frac\mu{m}\right)\right] + \cV_f \,,
\ee
where $\cV_f$ is finite as $d \to 4$. The constant $\cC$ is given in terms of the $s_i$ by
\be \label{eq:Cform}
 \cC := \frac{s_{-1}}{6} (m r)^6 - \frac{s_0}{2}(m r)^4 + s_1 (mr)^2 - s_2 \,.
\ee
The coefficients $s_i$ are functions of the bulk flux quantum, $\cN=\pm1$, the defect angles, $\alpha_b$, and the brane fluxes, $\Phi_b$.

\subsection*{Bulk divergences}

Because the wavelengths of interest are much shorter than the extra-dimensional size, divergences are instead absorbed into counter-terms in both the 6D bulk and 4D brane actions. Refs.~\cite{AccSUSY} show how to disentangle which bulk and brane interactions absorb the divergences found in eq.~\pref{eq:Vinfty}. However, for our purposes, it is sufficient to notice that the Gilkey coefficients $s_i$ decompose into a bulk and brane part as follows:
\be \label{sdecoup}
s_i = f(\alpha_+,\alpha_-) \, s_i^{\rm sph} + \sum_b\delta s_{ib} \,,
\ee
where the multiplying factor $f(\alpha_+,\alpha_-)$ reflects the change in volume due to the presence of the brane sources\footnote{In the case of non-identical branes, $f$ is
\be \label{falphapmdef}
f(\alpha_+,\alpha_-) = \frac{2\,(\alpha_+\alpha_-)^{3/4}}{\sqrt{\alpha_+} + \sqrt{\alpha_-}} \,.
\ee
}. The specific form of $f$ is not needed since the bulk contributions to the Gilkey coefficients are independent of the boundary conditions, and so are guaranteed to cancel --- when summed over a multiplet --- as they do in the Salam-Sezgin case \cite{RicciFlatUV}.
Physically, this is because the bulk counterterms capture the effects of very short-wavelength modes, which don't extend far enough through the extra dimensions to `know' about conditions imposed at the boundaries.

\subsection*{Brane divergences}

In a similar vein as the previous argument, since the brane corrections $\delta s_{ib}$ are capturing the effects of short-wavelength modes at the brane, they depend only on the local properties of each brane, and are insensitive to the properties of other distant branes. As such, their form as derived in \cite{AccSUSY} for the case of the rugby ball is valid for the case of non-identical branes as well. Therefore, there is no need to re-derive these Gilkey-de Witt coefficients. In subsection \ref{subsec:GdWcoeffs}, we recap the values of the $\delta s_i$'s for the field content of various supermultiplets.

\subsection*{Brane back-reaction}

Finally, we find some additional simplification as a result of back-reaction. Since the $\delta s_i$'s are renormalized by changing the brane couplings, the bulk geometry will back-react due to these 1-loop corrections. In \cite{LargeDims}, the back-reacted vacuum energy is shown to be
\be
\rho^{(\ssB\ssR)}_\ssV = \frac12 \sum_b\frac{\pd \cL_{b\ssR}}{\pd\phi} \,,
\ee
where $\cL_{b\ssR}$ is the renormalized brane lagrangian density: $S_b = \int \!\exd^4x \sqrt{-g} \,\cL_{b\ssR}$. Therefore, only contributions from massive multiplets can contribute to the final result since they are the only ones which can grow dilaton-dependence in the brane couplings:
\be \label{rhoVBR}
\rho^{(\ssB\ssR)}_\ssV = \frac{C}{(4\pi r^2)^2} \,,\quad {\rm where}\quad C = \sum_b \left[\frac{\delta s_{0b}}2 \left(\frac{\kappa M}{2\gR}\right)^4 - \frac{\delta s_{1b}}2 \left(\frac{\kappa M}{2\gR}\right)^2 \right] \ln \left(\frac{M_g}M\right) \,.
\ee
(In the above, $\delta s_{-1}$ does not appear because $s_{-1}$ is renormalized entirely by the bulk potential.) This back-reacted result is then added to the finite part of the 1-loop effective potential to determine the net 1-loop vacuum energy:
\be
\rho_\ssV = \rho_\ssV^{(\ssB\ssR)} + \cV_{f} = \frac{C+C_f}{(4\pi r^2)^2} \,.
\ee
 Although we do not compute $C_f$, we expect it to vanish in the supersymmetric case, and for its size to be $\cO(1)$ given previously computations on odd-dimensional spheres \cite{OrdonezRubin} (although graviton contributions may be enhanced). However, since several bulk fields are expected for anomaly cancellation \cite{6Danomcancel}, its contribution could be enhanced; its exact value remains to be checked. For a sufficiently large value of $\left(\tfrac{\kappa M}{2\gR}\right)$, we are guaranteed that $C$ is the dominant contribution to the vacuum energy.

\subsection{Assembly of Gilkey-de Witt coefficients}
\label{subsec:GdWcoeffs}

Since we argue in the previous subsection that the brane divergences are no different in the more general case of non-identical branes, as compared to the ones computed for the rugby ball, this subsection simply recaps the various brane Gilkey-de Witt coefficients denoted by $\delta s_i$, as found in \cite{AccSUSY}. This is done first for individual fields in a matter multiplet, and then assembled to give the desired result. Since only $\delta s_{0b}$ and $\delta s_{1b}$ appear in eq.~\pref{rhoVBR}, we track only these brane divergences here. (For convenience, we drop the subscript $b$ on the $\delta s_i$'s in this subsection.)

The bosonic sector of the 6D hypermultiplet is composed of four hyperscalars; they form a specific quaternionic potential, as dictated by supersymmetry. Each hyperscalar has the following Gilkey-de Witt coefficients:
\bea
\delta s_0^{\rm hs} &=& \frac1\omega\left( \frac{\omega^2-1}{12} - \frac{\omega^2}2 \, F(|\Phi|) \right) \\
\delta s_1^{\rm hs} &=& \frac1\omega \left( -\frac{\omega^2-1}{72} + \frac{\omega^4-1}{360} + \frac{\omega^2}{12} \, F(|\Phi|) - \frac{\omega^4}{12} \, F^2(|\Phi|) \right) \mp \frac{\omega^2}{12} \, \Phi \, G(|\Phi|)
\eea
where
\be
\omega:=1/\alpha \,,\quad F(x):= x(1-x) \,,\quad G(x) := (1-x)(1-2x) \,.
\ee
The uncharged spin-1/2 (6D) Weyl hyperino has
\bea
\delta s_0^{\rm f_0} &=& \frac1\omega\left( \frac{\omega^2-1}6  \right) \\
\delta s_1^{\rm f_0} &=& \frac1\omega \left( \frac{\omega^2-1}{72} + \frac{7(\omega^4-1)}{720} \right) \,.
\eea
%

The gauge multiplet contains a charged spin-1/2 Weyl fermion --- the gaugino --- whose Gilkey de-Witt coefficients are
\bea
\delta s_0^{\rm f} &=& \frac1\omega\left( -\frac{\omega^2-1}3 + \omega^2 \, \sum_{\sigma=\pm1} F(|\Phi_{\rm f\sigma}|) \right) \\
\delta s_1^{\rm f} &=& \frac1\omega \left( \frac{\omega^2-1}{18} - \frac{\omega^4-1}{90} - \frac{\omega^2}{6} \sum_{\sigma=\pm1} (1\mp3\sigma) F(|\Phi_{\rm f\sigma}|) + \frac{\omega^4}6 \sum_{\sigma=\pm1} F^2(|\Phi_{\rm f\sigma}|)  \right)   \nn\\
&&\quad\quad+ \frac{\omega^2}{6} \!\sum_{\sigma=\pm1} (\pm1-\sigma)  \Phi_{\rm f\sigma} \, G(|\Phi_{\rm f\sigma}|) \,,
\eea
where $\sigma = +1$ ($-1$) denotes positive (negative) helicity, and
\be
\Phi_{\rm f\sigma} := \Phi - \frac\sigma 2 (1-\alpha) \,.
\ee
The (uncharged) spin-1 gauge field has the following coefficients:
\bea
\delta s_0^{\rm gf} &=& \frac1\omega\left( -(\omega-1) +\frac{\omega^2-1}3 \right) \\
\delta s_1^{\rm gf} &=& \frac1\omega \left( \frac{\omega^2-1}{9} + \frac{\omega^4-1}{90} \right)  \,.
\eea
%

In the supersymmetric case, the flux is related to the defect angle in the following way:
\be \label{Phifluxval}
\Phi = \Phi_{\rm s} := \pm \frac12 (1-\alpha) \,.
\ee
Therefore, we can readily check that these Gilkey-de Witt coefficients cancel in the supersymmetric case. Specializing to the flux in eq.~\pref{Phifluxval}, the supersymmetric values for the charged coefficients are
\begin{gather}
\delta s_0^{\rm hs} \Big|_{\Phi=\Phi_{\rm s}} = -\frac{\omega^2-1}{24\,\omega} \,,\quad \delta s_0^{\rm f} \Big|_{\Phi=\Phi_{\rm s}} = \frac1\omega \left(\omega-1 - \frac{\omega^2-1}{3} \right) \,,\\
\delta s_1^{\rm hs} \Big|_{\Phi=\Phi_{\rm s}} = \frac1\omega\left( -\frac{\omega^2-1}{288} - \frac{7(\omega^4-1)}{2880} \right) \,,\quad \delta s_1^{\rm f} \Big|_{\Phi=\Phi_{\rm s}} = \frac1\omega \left(-\frac{\omega^2-1}{9} - \frac{\omega^4-1}{90} \right) \,.
\end{gather}
From these, we see that --- once summed over an entire multiplet --- the combinations $\delta s_i^{\rm hm}:= 4\,\delta s_i^{\rm hs} + \delta s_i^{f_0}$ and $\delta s_i^{\rm gm}:=\delta s_i^{\rm gf} + \delta s_i^{\rm f}$ (from an entire hypermultiplet or gauge multiplet, respectively) vanish in the supersymmetric case, as was found previously in \cite{AccSUSY}.

Massive multiplets can also exist in six dimensions; they contribute to brane divergences through the combination
\be
\delta s_i^{\rm mm} := \delta s_i^{\rm hm} + \delta s_i^{\rm gm}
\ee
with the understanding that one of the hyperscalars in the hypermultiplet is `eaten' by the gauge field in the gauge multiplet.

In this combination, there is some partial cancellation that take place; for convenience, let's write the result with a flux specified in units of the supersymmetric one:
\be
\kay:=\frac{\Phi}{\,\,\Phi_{\rm s}} \,.
\ee
These Gilkey-de Witt coefficients are
\bea
\delta s_0^{\rm mm} &=&  (\omega-1) \times \left\{
\begin{array}{cc}
1-|\kay| \,,\quad& |\kay| \leq 1 \\
0  \,,\quad& |\kay|\geq 1
\end{array}
\right.  \\
\delta s_1^{\rm mm} &=& \frac{(\omega^2-1)}{8\,\omega} \times\left\{
\begin{array}{lc}
\frac13\,\omega^2+1  &   |\kay| \leq 1\\
\qquad- |\kay| \left(\frac{|\kay|(\omega-1)+2\sigma_k}{\omega+1}\right) \Big[ \omega^2\Big(1-\frac23|\kay|\Big) +\frac23 \, \omega\Big(|\kay|-\sigma_k\Big)+1\Big]  \,,
\\
\Big[ |\kay|-\sigma_\kay - (|\kay|-1) \omega\Big]^2 \,, & |\kay| \geq 1
\end{array}
\right. \label{ds1mmgen}
\eea
where $\sigma_\kay := \kay/|\kay|$. To demonstrate more clearly the behaviour of these functions, we plot them for a fiducial value of $\omega=1.2$ in figure \ref{siplots}. Their general features are:
\begin{itemize}
\item they are both non-negative;
\item the highest power of $\Phi$ in $\delta s_i$ is $(2i+1)$ when $|\Phi|\leq |\Phi_{\rm s}|$ and $(2i)$ when $|\Phi|\geq |\Phi_{\rm s}|$;
\item the maximum values are $(\omega-1)$ and $(\omega^2-1)/(2\omega)$, respectively.
\end{itemize}
%
\FIGURE[h!]{
  \centering
\begin{tabular}{cc}
	\includegraphics[width=0.45\textwidth]{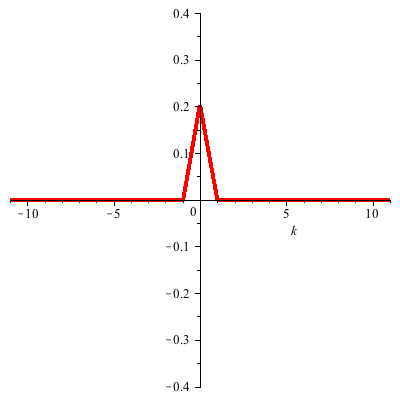} &
	\includegraphics[width=0.45\textwidth]{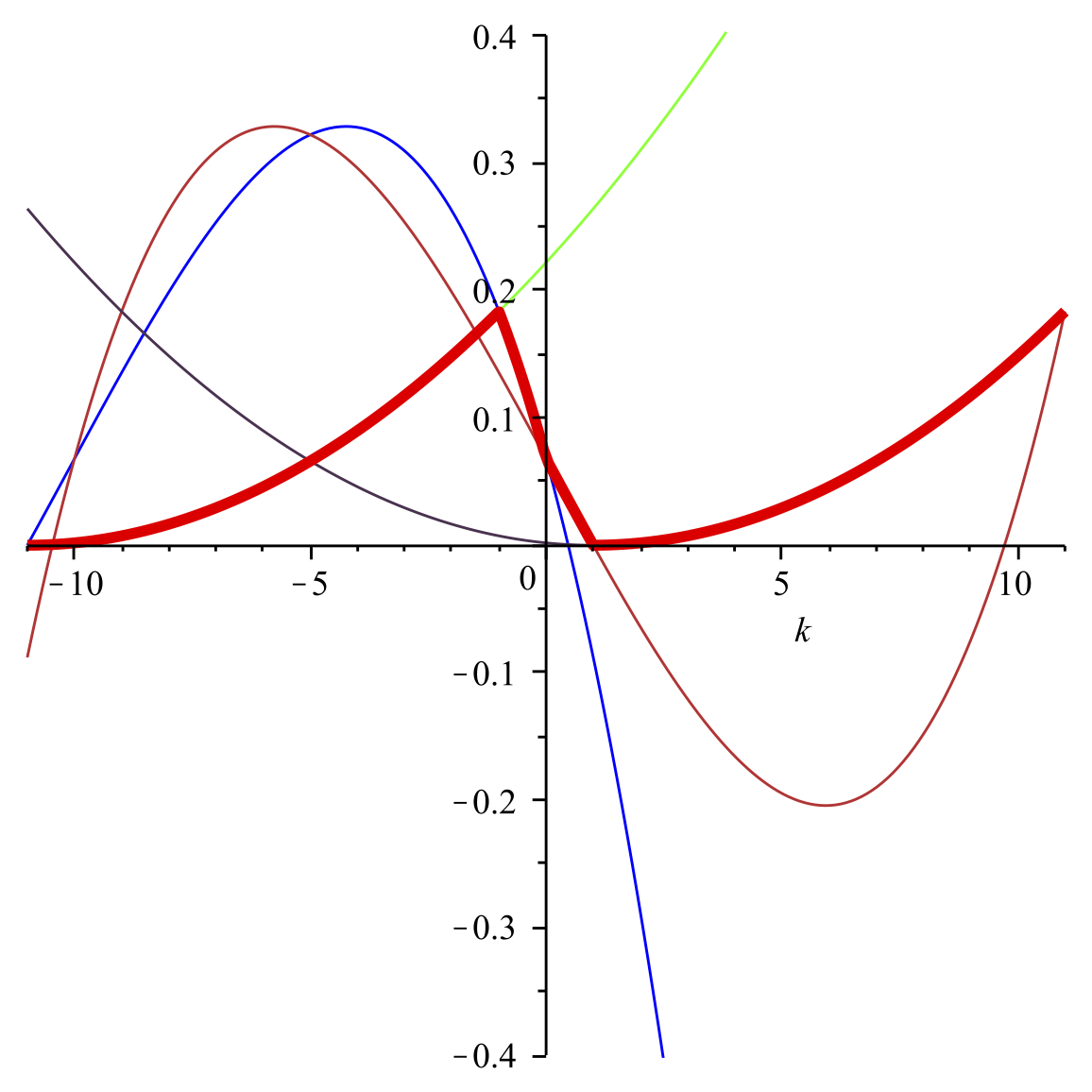}
\end{tabular}
    \caption{Plots of $\delta s_0$ (left) and $\delta s_1$ (right) for the fiducial value $\omega=1.2$, as a function of the flux in units of the supersymmetric one: $k=\Phi/\Phi_s$. In the second plot, Taylor expansions made in each disjointed regime of eq.~\pref{ds1mmgen} are overlaid; the function itself is in thick red (colour online).}
\label{siplots}
}

\subsection{Estimation of bounds}

In this subsection, we begin by performing a worst-case-scenario estimate of the types of expected bounds on the gravity scale and the extra-dimensional size. This estimate will appear bleak, but a more realistic analysis shows that one can simultaneously obtain the observed dark energy while avoiding tension with the known experimental constraints.

\subsection*{Worst-case scenario}

An upper limit on the kinds of bounds we expect to get can be obtained from considering $\delta s_0$ alone. Eq.~\pref{rhoVBR} tells us that the back-reacted vacuum energy is given by
\be \label{1loopvac}
\rho_\ssV^{(\ssB\ssR)} = \frac1{2(4\pi r^2)^2} \sum_b \left[ \delta s_{0b} \left(\frac{\kappa M}{2 \gR}\right)^4- \delta s_{1b} \left(\frac{\kappa M}{2\gR}\right)^2 \right] \ln \left(\frac{M_g}{M}\right) \,.
\ee
However, if this contribution is to dominate over the finite part of $\Vone$, then we expect
\be
\left(\frac{\kappa M}{2\gR}\right)^2 \gg 1 \,.
\ee
Therefore---since the maximum values of $\delta s_0$ and $\delta s_1$ are roughly the same size for small deviations from $\alpha=1$---we can estimate the maximum value of eq.~\pref{1loopvac} by taking $\delta s_{0b} \simeq (1-\alpha)$:
\be
\rho_\ssV \simeq \frac{(1-\alpha)}{(4\pi r^2)^2} \left(\frac{\kappa M}{2 \gR}\right)^4 \ln \left(\frac{M_g}{M}\right) \quad ({\rm worst\,case})\,.
\ee

To get a sense of the bounds that this type of expression might predict, let's take
\be
\gR = \left(0.01\,\tilde g\right) M_g^{-1} \,,\quad M = 0.1 \, M_g
\ee
(where $M_g = \kappa^{-1/2}$ as before). This gives
\be
\left(\frac{\kappa M}{2 \gR}\right)^2 = \frac{25}{\tilde g^2}\,\, \gg1 \quad\leftrightarrow\quad \tilde g \leq 1
\ee
and so, for small deviations from $\alpha=1$, we have
\be
\rho_\ssV \simeq \rho_\ssV^{\rm obs.} \left[\frac{T}{(5 \,{\rm TeV})^4}\right] \left(\frac{2.94\times 10^{15}}{\tilde g \, M_g r} \right)^4
\ee
where $\rho_\ssV^{\rm obs.} :=(2.3\times 10^{-3} \,{\rm eV})^4$. Since the radius and the gravity scale are related by
\be
\frac1{\kappa_4^2} \simeq \frac{4\pi r^2}{\kappa^2} = (2.4 \times 10^{18} \,{\rm GeV})^2 \,,
\ee
we obtain the following estimates:
\bea
M_g &=& (233 \, {\rm GeV})\, \tilde g \left(\frac{\rho_\ssV}{\rho_\ssV^{\rm obs.}}\right)^{1/4} \left(\frac{5 \,{\rm TeV}}{T}\right)^{1/4} \\
r &=& \frac{2.48 \, {\rm mm}}{\tilde g^2} \left(\frac{\rho_\ssV^{\rm obs.}}{\rho_\ssV}\right)^{1/2} \left(\frac{T}{5 \,{\rm TeV}}\right)^{1/2}  \,.
\eea
In each case, we find disagreement with observational bounds when evaluated at the fidicual values. In fact, in the case of $M_g$, we are no longer in the perturbative regime since $1-\alpha$ is no longer a small quantity. However, since these estimates are all made in the worst-case scenario, let's next consider a more optimistic one in which we can better exploit the benefits of a supersymmetric background.

\subsection*{Improved scenario: perturbing about a supersymmetric configuration}

In this scenario, we consider a supersymmetric rugby-ball configuration, which is perturbed due to a difference in tensions arising from a brane-particle loop. We begin by quoting the leading-order term in the large-mass limit, $\delta s_0$, as found in the previous subsection:
\be \label{ds0totdef}
\delta s_0 = \delta s_{0+} + \delta s_{0-} \,,\quad \delta s_{0b} = \frac2{\alpha_b} \times \left\{
\begin{array}{cc}
|\Phi_{{\rm s}b}|- |\Phi_b| \,,& |\Phi_b| \leq |\Phi_{sb}| \\
0 \,,& |\Phi_b| \geq |\Phi_{sb}|
\end{array}
\right.\,.
\ee
Such a $\delta s_0$ is plotted in Figure \ref{ds02branes}, as a function of the flux and defect angle differences. (Recall: the sum of fluxes is fixed by flux quantization.)
%
\FIGURE[h!]{
  \centering
    \includegraphics[width=0.6\textwidth]{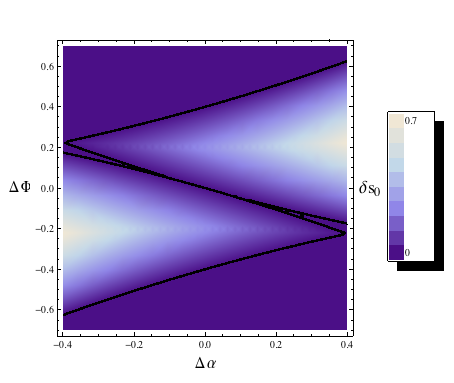}
    \caption{Plot of $\delta s_0$ for the fiducial value $\ol\alpha := \frac{\alpha_1+\alpha_2}2=0.8$, as a function of $\Delta\alpha:=\alpha_+ - \alpha_-$ and $\Delta\Phi:= \Phi_+ - \Phi_-$. The black line indicates the discontinuity at $\delta s_0 =0$ (colour online).}
\label{ds02branes}
}
%
Let's consider perturbing around a rugby-ball background with identical branes:
\be
\alpha_b = \alpha_0 + \delta \alpha_b \,,\quad \Phi_b = \pm\frac12(1-\alpha_0) + \delta \Phi_b \,.
\ee
Furthermore, let's assume that we are integrating out a brane particle at the north brane only:
\be
\delta T_+ = (-)^{1-\ssF}\frac{m^4}{2(4\pi)^2}\ln\left(\frac{M_g}{M}\right) \,,\quad \delta T_- = 0 \,.
\ee
(We find that, counterintuitively, the positive sign corresponds to fermionic loop upon integrating the brane particle beta function down from the gravity scale.) After some algebra (see Appendix \ref{app:pertmatch} for details), we find at leading order
\be
\delta \alpha_+ = -\frac34 \,\frac{\kappa^2 \delta T_+}{2\pi} \,,\quad \delta \alpha_- = \frac14 \,\frac{\kappa^2 \delta T_+}{2\pi} \,,\quad \delta \Phi_+ = \delta \Phi_- = \pm\frac18 \,\frac{\kappa^2 \delta T_+}{2\pi} \,.
\ee
When $\delta T_+ <0$ ({\it i.e.}~for bosonic loops) we have $|\Phi_b| < |\Phi_{{\rm s}b}|$ for both branes, moving us up the right-hand side of two triangles like the one in Figure \ref{siplots}. Therefore, at leading order in eq.~\pref{ds0totdef} we find
\be
\delta s_{0}  \simeq \frac{\kappa^2 |\delta T_+|}{4\pi} = \frac{\kappa^2 \,m^4}{2(4\pi)^3} \ln\left(\frac{M_g}{M}\right)\,.
\ee
Repeating our previous estimate in this more realistic case, we find that
\bea
\rho_\ssV &\simeq& \frac{\kappa^2\, m^4}{4(4\pi)^5 r^4} \left(\frac{\kappa M}{2 \gR}\right)^4 \left[\ln \left(\frac{M_g}{M}\right) \right]^2 \nn\\
&\simeq& (2.3\times 10^{-3} \,{\rm eV})^4 \left(\frac{m}{173 \,{\rm GeV}}\right)^4 \left(\frac{1.71\times 10^{13}}{\tilde g \, M_g r}  \right)^4
\eea
This --- together with $M_p = \sqrt{4\pi} M_g^2r$ --- yields the following estimates:
\bea
M_g &=& (40 \,{\rm TeV})\,\tilde g \,\left(\frac{\rho_\ssV}{\rho_\ssV^{\rm obs.}}\right)^{1/4} \left(\frac{173 \,{\rm GeV}}{m}\right) \\
r &=& \frac{0.083 \,\mu{\rm m}}{\tilde g^2} \left(\frac{\rho_\ssV^{\rm obs.}}{\rho_\ssV}\right)^{1/2} \left(\frac{m}{173 \,{\rm GeV}}\right)^{2}\,.
\eea
(In the above, we use a fidicual value of $m = 173$ GeV for concreteness, despite the assumption of a bosonic loop correction to the tension.)

To get a sense of how much freedom is allowed by these bounds, consider the plots in Figure \ref{boundplots}.
%
\FIGURE[h!]{
\centering
\begin{tabular}{cc}
	\hspace{-0.2in}
	\includegraphics[width=0.5\textwidth]{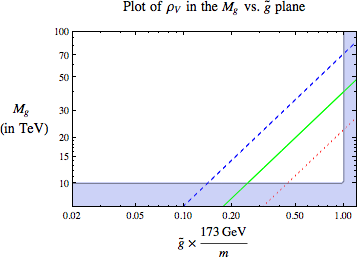} &\hspace{-0.1in}
   \includegraphics[width=0.5\textwidth]{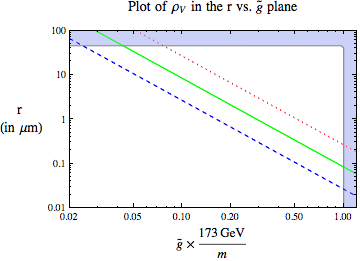}
\end{tabular}
\caption{Log plots of the vacuum energy in the $M_g$ vs.~$\tilde g$ and $r$ vs.~$\tilde g$ planes, respectively, for $\rho_\ssV = \rho_\ssV^{\rm obs.}$ (solid green), $\rho_\ssV = 10\times\rho_\ssV^{\rm obs.}$ (blue dashes), and $\rho_\ssV = 0.1\times\rho_\ssV^{\rm obs.}$ (red dots). The greyed regions are excluded (colour online).}
\label{boundplots}
}
%
Therein, we find that there is considerable parameter space available to obtain a vacuum energy which is comparable to the observed value. In particular, the region of parameter space accessible at the LHC would predict an extra-dimensional size in the range 0.1--1 $\mu$m.

\section{Discussion}
\label{sec:Discussion}

In this paper we have considered the implications of recent calculations of the vacuum energy in the scenario of large supersymmetric extra dimensions. We generalize earlier results in rugby ball geometries to extra dimensions that have a more general geometry, including warping. Our results confirm the expectation that the vacuum energy in these models is robustly set by the KK scale.

We identify the underlying symmetries that protect the vacuum energy at scales larger than the KK scale. The crucial ingredient is that the scale of supersymmetry breaking in the bulk is -- surprisingly-- not set by the mass splittings of standard model particles with their superpartners. Instead, the bulk SUSY breaking scale is set by gradients in the background fields that differ from the supersymmetric choice, and those are all set by the scale of the extra dimensions. The reason for this is that the BPS-like condition on the branes that relates its flux to its tension, can be satisfied exactly locally at each brane separately. It is only when those local conditions are mutually inconsistent due to global considerations like flux quantization that the supersymmetry in the bulk is broken.

Quantitatively, we show that there is significant parameter space for which these models are consistent with the current state of the art of detecting extra dimensions. The next generation of collider and inverse-square-law experiments will be capable of determining the validity of such an extra-dimensional origin for the observed vacuum energy.

\section*{Acknowledgements}

We are grateful to Arthur Hebecker for asking the question that precipitated a more detailed computation of the vacuum energy density in these models, and we thank Itay Yavin for many useful discussions. The Abdus Salam International Centre for Theoretical Physics (ICTP) kindly supported and hosted various combinations of us while part of this work was done. Our research was supported in part by funds from the Natural Sciences and Engineering Research Council (NSERC) of Canada. Research at the Perimeter Institute is supported in part by the Government of Canada through Industry Canada, and by the Province of Ontario through the Ministry of Research and Information (MRI).

\appendix

\section{Some properties of the independent-brane solutions}

This appendix records some of the properties of the geometry discussed in \S2\ that describes the bulk sourced by two non-identical branes.

\subsection{Relation to GGP coordinates}
\label{app:GGPcoords}

This section derives the form used for the metric, eq.~\pref{eq:GGPmetric}, by performing a coordinate change from the solution of ref.~\cite{GGP}, which derives the solution using the ansatz
\begin{gather} \label{warpedmetric}
 \exd s^2 = W^2(\eta) \, \exd s_4^2 + a^2(\eta)\, e^{-\phi_0} \Big( W^8(\eta)\, \exd\eta^2 + \exd \varphi^2 \Big) \nn\\
 F_{\mu\ssN} =0 \,,\quad F_{mn} = f(\eta) \,\epsilon_{mn} \,,\quad \phi = \phi(\eta) \,.
\end{gather}
In these coordinates, the Ricci tensor has the following non-vanishing components,
\bea
 \cR_{\mu\nu} &=& \frac1{a^2 e^{-\phi_0} W^8} \left[ \frac{\pd_\eta^2 W}{W} - \left( \frac{\pd_\eta W}{W} \right)^2 \right] g_{\mu\nu} \nn\\
 \cR_{\eta\eta} &=& \frac{\pd_\eta^2 a}a - \left( \frac{\pd_\eta a}{a} \right)^2 + 4\, \frac{\pd_\eta^2 W}W   - 8 \left( \frac{\pd_\eta a}a \right) \frac{\pd_\eta W}W - 16 \left( \frac{\pd_\eta W}{W} \right)^2\\
 \cR_{\varphi\varphi} &=& \frac1{W^8} \left[ \frac{\pd_\eta^2 a}a - \left( \frac{\pd_\eta a}{a}\right)^2 \right]  \,, \nn
\eea
and the equations of motion can be integrated to yield the following solution \cite{GGP}:
\bea
 W^4(\eta) &=& \frac{\cosh[\lambda(\eta-\eta_1)]}{\cosh[\lambda(\eta-\eta_2)]} \\
 a^4(\eta) &=& \frac{a_0^4}{\cosh^{3}[\lambda(\eta-\eta_1)] \cosh[\lambda(\eta-\eta_2)]} \qquad\\
 F_{\eta\varphi} &=& \left( \frac{2\gR^2}{\kappa^2} \right)   \frac{a^2(\eta)}{W^2(\eta)} =  \pm\frac{\lambda^2}{2 \cosh^2[\lambda(\eta-\eta_1)]}  \\
 e^{\phi(\eta)} &=& \frac{e^{\phi_0}}{W^2(\eta)} \,.
\eea
Here, $a_0 := \lambda \kappa / 2 \gR$ and the three independent integration constants are $\lambda$, $\Delta\eta:=\eta_2-\eta_1$, and $\phi_0$.

To obtain the form used in \S2, first define
\be
 r(\eta) = \frac\kappa{2\gR}\, e^{-\phi(\eta)/2} = r_0\, W(\eta) \,,\quad{\rm with}\quad r_0 := \frac\kappa{2\gR}\, e^{-\phi_0/2} \,.
\ee
The coordinate $\theta(\eta)$ is obtained by requiring $r(\eta) \,\exd \theta = a(\eta) e^{-\phi_0/2} \,W^4(\eta)\, \exd \eta$, which gives
\be
 \exd \theta = \frac{\lambda \,\exd \eta}{\cosh[\lambda(\eta-\eta_2)]} \,.
\ee
Integrating from $\theta=0$ (corresponding to $\eta\to-\infty$) yields the three equivalent forms
\be
 \theta(\eta) =2 \arctan\!\big[ e^{\lambda(\eta-\eta_2)}\big] \;\;{\rm\bf or}\;\; e^{\lambda\eta(\theta)} = e^{\lambda\eta_2} \tan\left(\frac\theta 2\right) \;\;{\rm \bf or}\;\; \sin\theta(\eta) = \frac1{\cosh[\lambda(\eta-\eta_2)]} \,.
\ee
From these we see that $\eta\to+\infty$ corresponds to $\theta=\pi$.

With this relation in tow we can find the connection between the integration constants used here and those appearing in \S2. Evaluating the warp factor gives
\bea
 W^4(\theta) := \frac{\cosh[\lambda(\eta(\theta)-\eta_1)]}{\cosh[\lambda(\eta(\theta)-\eta_2)]} &=& \frac{e^{\lambda\Delta\eta} \tan(\theta/2) + e^{-\lambda\Delta\eta} \cot(\theta/2)}{\tan(\theta/2) + \cot(\theta/2)} \\
 &=& e^{\lambda\Delta\eta} \sin^2 \frac{\theta}2 + e^{-\lambda\Delta\eta} \cos^2 \frac{\theta}2  \,, \nn
\eea
and so $\xi = \lambda \Delta \eta$.

Similarly
\be
 \frac{a(\eta) \, e^{-\phi_0/2}}{r(\eta)} = \frac\lambda{\cosh[\lambda(\eta-\eta_1)]} = \frac\lambda{W^4 \cosh[\lambda(\eta-\eta_2)]} \,,
\ee
and so
\be
 g_{\varphi\varphi} := a^2 e^{-\phi_0} = \left(\frac\lambda{W^4}\right)^2 r^2 \sin^2\theta\,,
\ee
leading to the line element of \S2:
\bea
 \exd s^2 &=& W^2(\theta) \, \exd s_4^2 + r^2(\theta) \Big(\exd\theta^2 + \alpha^2(\theta) \sin^2\theta \,\exd\varphi^2 \Big) \\
 &=&W^2(\theta) \Big[ \exd s_4^2 + r_0^2 \Big(\exd\theta^2 + \alpha^2(\theta) \sin^2\theta \,\exd\varphi^2 \Big)\Big] \,,
\eea
where $r_0$ and $\alpha(\theta)$ are as defined there. The gauge field is similarly given by
\be
 \cF_{\theta\varphi} = \frac{\exd\eta}{\exd\theta} \, \cF_{\eta\varphi} = \pm \frac{\lambda \,\sin\theta}{2 \, W^8(\theta)} = \pm\frac1{2 \, r^2(\theta)} \, \frac{\epsilon_{\theta\varphi}}{W^4(\theta)} \,,
\ee
as in the main text.

\subsection{Brane-bulk matching conditions}
\label{app:pertmatch}

In this section we record how the bulk integration constants are related to the brane positions by the brane-bulk matching conditions, including both tension and brane-localized flux at each brane. The three integration constants in the bulk solution can be traded for these four brane properties because flux quantization imposes one relation between them.

As shown in \cite{HiCoDBCs}, in the presence of brane-localized flux the graviton boundary conditions at each codimension-2 brane require
\be
 1-\alpha_b = \frac{\kappa^2 L_b}{2\pi} \,,
\ee
where
\be
 L_b := T_b - \frac{\cA_b}{2\gR^2} \, \epsilon^{mn} F_{mn}
 = T_b \mp \frac{4\pi \Phi_b}{\kappa^2 W_b^4} \,,
\ee
and the last equality evaluates the result on the background solution, using
\bea \label{PhiAdef}
 \Phi_b: &=& \frac{\cA_b e^{\phi_b}}{2\pi}= \frac{\cA_b e^{\phi_0}}{2\pi W_b^2} \\
 \hbox{and} \quad W_b^4 &=& e^{-b\lambda\Delta\eta} = \frac{\sqrt{\alpha_+\alpha_-}}{\alpha_b} \,.
\eea
Combining these expressions gives the result
\be
 1-\alpha_b  = \frac{\kappa^2 T_b}{2\pi} \mp \frac{2\Phi_b}{W_b^4} \,,
\ee
which is to be solved for $\alpha_b$, say, keeping in mind that $\alpha_b$ also appears implicitly in the expressions for $W_b$ on the right-hand side.

To proceed further it is useful to assume that the coefficients $\cA_b$ are the same for both branes: $\cA_+ = \cA_- := \cA$. Besides simplifying later formulae, this is also the near-supersymmetric situation of interest in the main text. In this situation we imagine starting with identical branes (the supersymmetric case), and add non-supersymmetric on-brane particles (like the Standard Model) on one of them, without coupling them to the brane-localized flux. In this case loops of brane particles can generate a tension difference, $T_+ \ne T_-$, but the quantities $\cA_\pm$ remain equal. However, it is important to recognize that this does {\em not} also imply the physical flux, $\Phi_\pm$, localized on the two branes need be identical, because of the $W_b$-dependence of $\Phi_b$.

With this assumption the flux quantization condition fixing the zero mode, $\phi_0$, is
\be
 \sum_b \Phi_b = \frac{\cA \, e^{\phi_0}}{2\pi} \, \frac{\sqrt{\alpha_+}+\sqrt{\alpha_-}}{\big(\alpha_+\alpha_-\big)^{1/4}}=\pm \big(1-\sqrt{\alpha_+\alpha_-}\big) \,.
\ee
Solving for $e^{\phi_0}$ and eliminating it from eq.~\pref{PhiAdef}, we find the a relation between $\Phi_b$ and the defect angles, $\alpha_b$:
\be
 \Phi_b = \frac{\pm \big(1-\sqrt{\alpha_+\alpha_-}\big)}{\sqrt{\alpha_+} + \sqrt{\alpha_-}} \, \sqrt{\alpha_b}  \,.
\ee
Therefore, using this in the graviton boundary conditions, gives the following non-linear expression to be solved to obtain $\alpha_b$ as a function of brane tension:
\be\label{AppDefectAngles}
 1-\alpha_b = \frac{\kappa^2 T_b}{2\pi} - \frac{2\big(1-\sqrt{\alpha_+\alpha_-}\big)}{\big(\sqrt{\alpha_+}+\sqrt{\alpha_-}\big)\sqrt{\alpha_+\alpha_-}}\,\alpha_b^{3/2} \,.
\ee
Notice that the function in front of the factor $\alpha_b^{3/2}$ is symmetric in the interchange $\alpha_+ \leftrightarrow \alpha_-$.

Eq.~\pref{AppDefectAngles} can be solved explicitly in the semiclassical limit, for which the combinations $\kappa^2 T_\pm/2\pi$ are both small. Working to first order in these quantities gives the results
\ba \label{aeq:alphavsTpm}
 \alpha_+ &=& 1 - \frac{3\kappa^2 T_+}{8\pi} + \frac{\kappa^2 T_-}{8\pi}\,,\nn\\
 \alpha_- &=& 1 - \frac{3\kappa^2 T_-}{8\pi} + \frac{\kappa^2 T_+}{8\pi}\,.
\ea
Notice that the presence of brane-localized flux makes the local defect angle at each brane depend on the tensions at {\em both} branes. In the special case where $T_+ = T_- := T$, eq.~\pref{aeq:alphavsTpm} reduces to the rugby-ball result \cite{LargeDims}: $1-\alpha =  \kappa^2 T/4\pi$, which (because of the brane-localized flux) is half as large as the standard `pure-tension' expression \cite{Vilenkin}.

In the particular case where we start out with equal tensions, but on-brane loops perturb only the tension of the $+$ brane, so $T_- = T$ and $T_+ = T + \delta T$, the defect angles at {\em both} branes are perturbed, and we have
\be
 \delta \alpha_+ = - \frac{3\kappa^2\delta T}{8\pi} \,,\quad \delta \alpha_- = \frac{\kappa^2 \delta T}{8\pi}\,.
\ee

\end{document}